\documentclass[aps,preprint]{revtex4}%
\usepackage{amsfonts}
\usepackage{amsmath}
\usepackage{amssymb}
\usepackage{graphicx}%
\begin{document}
\title[ENTROPY PRINCIPLE FOR BLACKBODY RADIATION]{DERIVATION OF THE BLACKBODY RADIATION SPECTRUM FROM A NATURAL MAXIMUM-ENTROPY
PRINCIPLE INVOLVING CASIMIR ENERGIES AND ZERO-POINT RADIATION}
\author{Timothy H. Boyer}
\affiliation{Department of Physics, City College of the City University of New York, New
York, NY 10031}

\begin{abstract}
By numerical calculation, the Planck spectrum with zero-point radiation is
shown to satisfy \ a natural maximum-entropy principle whereas alternative
choices of spectra do not. \ Specifically, if we consider a set of
conducting-walled boxes, each with a partition placed at a different location
in the box, so that across the collection of boxes the partitions are
uniformly spaced across the volume, then the Planck spectrum corresponds to
that spectrum of random radiation (having constant energy $k_{B}T$ per normal
mode at low frequencies and zero-point energy $(1/2)h\omega$ per normal mode
at high frequencies) which gives maximum uniformity across the collection of
boxes for the radiation energy per box. \ Our analysis involves Casimir
energies and zero-point radiation which do not usually appear in thermodynamic
analyses. \ For simplicity, the analysis is presented for waves in one space
dimension. \ 

\end{abstract}
\maketitle

\bigskip\draft

\section{\noindent\textbf{INTRODUCTION}}

A thermodynamic system described by a known Lagrangian has well-defined
restrictions on its adiabatic curves following from the first law of
thermodynamics. \ However, the mechanical Lagrangian does not give the full
thermodynamic behavior because it does not determine the thermodynamic
entropy. \ Rather, some additional idea of order, of uniformity, is required
for the determination of the entropy function. \ The assumptions of
statistical mechanics regarding equally-probable units provide the idea of
uniformity needed to derive the entropy function. \ Sometimes the idea of
uniformity is regarded as immediately evident or natural. \ For example, in
the classical statistical mechanics of particles in a box, it is natural to
assume that the situation of maximum uniformity, maximum entropy, corresponds
to a uniform spatial distribution for particles in the box. \ In this article,
we propose that there is a similar natural choice for maximum entropy for
blackbody radiation in connection with Casimir energies and zero-point
radiation; this natural choice leads to the Planck spectrum.

Blackbody radiation is a fundamental thermodynamics system which holds a
special place in the history of physics as the beginning point of quantum
theory.[1] \ In the nineteenth century, Boltzmann used Maxwell's connection
between radiation pressure and energy density, together with the assumption
that the energy density depended upon temperature alone, to derive Stefan's
law connecting total energy density and temperature.[2] \ Wien used adiabatic
compression and Maxwell's theory to obtain the displacement law corresponding
to the condition on the adiabatic curves for a harmonic oscillator
thermodynamic system.[3] \ However, determination of the full blackbody
radiation spectrum seemed to confound classical analysis because no natural
entropy analysis seemed possible. \ Today, the blackbody spectrum is regarded
as comprehensible only in terms of energy quanta which are outside of
classical theory.

However, in connection with random radiation, twentieth century physics
contributed two important ideas, zero-point energy and Casimir forces, which
raise new possibilities for recognizing a natural entropy function for thermal
radiation. \ Zero-point energy is random energy which is present even at zero
temperature.[4] Thermodynamics allows the possibility of zero-point energy and
experimental evidence such as van der Waals forces requires its existence.[5]
\ Casimir forces and energies are those which arise due to the discrete,
classical normal modes structure of a system.[6] \ In total contrast to
particles, waves are influenced in the interior of a volume by the presence of
boundary conditions at the walls. \ Thus if a thin conducting partition is
introduced into a conducting-walled box, then the energy of the system is
changed due to the new boundary conditions at the conducting partition.

Casimir energies serve to couple total electromagnetic radiation energy to the
specific spectrum of random radiation. \ Thus if we consider a collection of
conducting-walled containers, each with a partition at a different location,
and each box having random radiation at the same temperature, then each of
these boxes will have a different (average) thermal energy. \ And different
assumed spectral distributions for thermal radiation will lead to different
distributions of energies among the partitioned boxes. \ Now maximum entropy
naturally suggests uniformity. \ It seems natural to suggest that maximum
entropy corresponds to the least variation of energy among the partitioned
boxes. \ We will find that, in the presence of zero-point radiation to prevent
an "ultra-violet catastrophe," this natural entropy idea leads to the Planck
spectrum for thermal radiation.

\section{\noindent\noindent\textbf{THERMODYNAMICS OF WAVES IN A
ONE-DIMENSIONAL BOX}}

Although the calculations to be described below can be carried through for
electromagnetic waves in a three-dimensional box, we will consider a
thermodynamic wave system in one spatial dimension rather than in three
because the mathematics is distinctly simpler while the physical ideas are
unchanged.[7] \ Thus we can imagine one-dimensional thermodynamic wave systems
consisting of waves on a string, or of electromagnetic waves which are
required to move between two conducting walls with wave vectors $\mathbf{{k}}$
which are always perpendicular to the walls.

Systems satisfying the wave equation in a container with conducting walls can
be described in terms of normal modes of oscillation, each of which
corresponds to a harmonic oscillator system[8]\noindent\ with Lagrangian%
\begin{equation}
L(q_{\lambda},\dot{q}_{\lambda})=\Sigma_{\lambda}(1/2)(\dot{q}_{\lambda}%
^{2}-\omega_{n}^{2}q_{\lambda}^{2})
\end{equation}
where the $q_{\lambda}$ are the amplitudes of the normal modes. For waves in
one spatial dimension inside a box of length $L$, the normal modes can be
labeled by a single integer index $n$ where the associated frequency
$\omega_{n}$ is given by $\omega_{n}=cn\pi/L$, $n=1,2,3...$, where $c$ is the
speed of the waves. \ Wien's displacement law, which follows from the
application of the laws of thermodynamics to a harmonic oscillator system,
tells us that the energy $\mathcal{U}$ of a normal mode at frequency $\omega$
and temperature $T$ is given by
\begin{equation}
\mathcal{U}(\omega,T)=-\omega\phi^{\prime}(\omega/T)
\end{equation}
where $\phi^{\prime}$ is a function of the single variable $\omega/T$.[9] \ 

Wien's displacement law allows two limits which make the energy $\mathcal{U}$
independent of one of its variables. \ Thus if $\phi^{\prime}\rightarrow
const$ when $\omega/T>>1$, then $\mathcal{U}$ depends upon $\omega$ alone.
\ This corresponds to temperature-independent zero-point radiation
\begin{equation}
\mathcal{U\rightarrow U}_{zp}(\omega)=(1/2)\hbar\omega\text{ \ for \ }%
\omega/T>>1
\end{equation}
where the constant $\hbar$ must be chosen to have the value of Planck's
constant in order to fit with van der Waals forces. \ On the other hand, if
$\phi^{\prime}\rightarrow const/(\omega/T)$, when $\omega/T<<1$, then
$\mathcal{U}$ depends upon $T$ alone. \ This corresponds to the
\ Rayleigh-Jeans spectrum%
\begin{equation}
\mathcal{U\rightarrow U}_{RJ}(\omega)=k_{B}T\text{ \ \ for \ \ }\omega/T<<1
\end{equation}
which holds at low frequencies with $k_{B}$ as Boltzmann's constant. \ 

Since we are not interested in numerical calculations of thermodynamic
quantities, we will use natural units[10] where $\hbar=1$ so that frequency is
measured in energy units, and $k_{B}=1$ so that temperature is measured in
energy units and entropy is a pure number,%
\begin{equation}
\mathcal{U}_{zp}(\omega)=(1/2)\omega\text{ \ \ and \ \ \ }\mathcal{U}%
_{RJ}(\omega)=T
\end{equation}

It is convenient to introduce the thermal energy $\mathcal{U}_{T}(\omega,T)$
of a mode of frequency $\omega$ as the mode energy above the zero-point
energy
\begin{equation}
\mathcal{U}_{T}(\omega,T)=\mathcal{U}(\omega,T)-\mathcal{U}_{zp}(\omega)
\end{equation}
Although the total energy $\mathcal{U}$ is related to forces, only the thermal
energy $\mathcal{U}_{T}$ influences the entropy.[11] \ 

The total thermal radiation energy $U_{T}$ in a box of length $L$ is given by
the sum over the thermal energies $\mathcal{U}_{T}$ of the modes of
frequencies $\omega_{n}=n\pi c/L$ for integer $n.$ \ We consider only the
thermal energy $\mathcal{U}_{T}$ since thermodynamics requires that $U_{T}$ is
a finite quantity when summed over all normal modes. \ In contrast, use of the
modes' total energies $\mathcal{U}$ or zero-point energies $\mathcal{U}_{zp}$
will give a divergence in the sum over (infinitely many) high-frequency modes.
In a one-dimensional box which is so large that the discrete sum over normal
modes can be replaced by an integral, we can use (2) to obtain the total
thermal energy $U_{T}(L,T)$ in the form
\[
U_{T}(L,T)=\Sigma_{n=1}^{\infty}\mathcal{U}_{T}(\frac{n\pi c}{L}%
,T)=\Sigma_{n=1}^{\infty}\frac{n\pi c}{L}\left[  \phi^{\prime}(\frac{n\pi
c}{LT})-\frac{1}{2}\right]
\]

\begin{equation}
\approx\int_{0}^{\infty}dn\frac{n\pi c}{L}\left[  \phi^{\prime}(\frac{n\pi
c}{LT})-\frac{1}{2}\right]  =\frac{L}{c\pi}T^{2}\int_{0}^{\infty}dzz\left[
\phi^{\prime}(z)-\frac{1}{2}\right]
\end{equation}
for one space dimension. This is just the Stefan-Boltzmann result appropriate
for one space dimension.[12] In the case of waves in three spatial dimensions,
the frequencies of the normal modes $\omega_{lmn}$ would be labeled by three
integer indices and the same procedure would lead to a $T^{4}$ temperature
dependence for a large container.

\section{\noindent\textbf{NORMAL MODE STRUCTURE AND CASIMIR FORCES}}

The Stefan-Boltzmann law in (7) gives the temperature dependence of the total
thermal energy but provides no information regarding the spectrum of thermal
radiation. Now in obtaining Eq.(7), we took the limit of a large box $L$ and
so replaced the sum over normal modes by an integral. However, by going to the
continuum limit, we lost the information which might be available in the
discrete spectrum of the normal modes. It was Casimir who saw the possibility
of new forces and energies linked to this discreteness of the classical normal
mode structure. The most famous example of such forces is the original Casimir
calculation[6] of the force between conducting parallel plates arising from
electromagnetic zero-point radiation. Casimir worked specifically with
zero-point fields; however, the idea is not limited to zero-point radiation.
Any spectrum of random classical radiation will lead to Casimir energies
associated with the discrete classical normal mode structure of a
container.[13] Indeed, every thermodynamic variable (energy, entropy, free
energy, force) will depend upon the normal modes structure.

It should be emphasized how totally different this classical wave situation is
from the classical particle situation of ideal gas particles in a box. Thus if
a box with reflecting walls is filled with ideal gas particles at temperature
$T$, then the introduction of a thin reflecting partition does not change the
system energy and does not involved any average force on the partition. In
total contrast, the introduction of a conducting partition into a
conducting-walled box of thermal radiation leads to a change in the normal
mode structure and hence both to position-dependent energy changes (Casimir
energies) and to average forces on the partition (Casimir forces). These
Casimir energies and forces will depend upon the precise spectrum of random
radiation and upon the precise location of the partition. In this article we
note that the Planck spectrum for thermal radiation equilibrium can be
obtained from a natural maximum-entropy principle for the Casimir energy
changes associated with the placement of partitions in boxes of radiation.

\section{\noindent\textbf{CHANGE IN CASIMIR ENERGY DUE TO A PARTITION}}

We now consider a one-dimensional box of length $L$ and calculate the change
of radiation energy $\Delta U(x,L,T)$ with position $x$ for a partition which
is located a distance $x$ from the left-hand end of the box, $0\leq x\leq L$.
The energy of each normal mode of frequency $\omega_{n}$ is given by
$\mathcal{U}(\omega_{n},T)$. The partition changes the normal mode frequencies
and so produces a position-dependent energy change $\Delta U(x,L,T)$ which is
a Casimir energy. We will calculate the Casimir energy $\Delta U(x,L,T)$ as
the change in the system energy when the partition is placed a distance $x$
from the left-hand wall compared to when the partition is placed at $x=L/2$ in
the middle of the box,
\[
\Delta U(x,L,T)=\left\{  U(x,T)+U(L-x),T\right\}  -\left\{
U(L/2,T)+U(L/2,T)\right\}
\]%
\begin{equation}
=\left\{  \Sigma_{n=1}^{\infty}\mathcal{U}\left(  \frac{cn\pi}{x},T\right)
+\Sigma_{n=1}^{\infty}\mathcal{U}\left(  \frac{cn\pi}{L-x},T\right)  \right\}
-2\Sigma_{n=1}^{\infty}\mathcal{U}\left(  \frac{cn\pi}{L/2},T\right)
\end{equation}

\section{\noindent\textbf{CASIMIR ENERGY FOR THE ZERO-POINT SPECTRUM}}

Equation (8) for the Casimir energy $\Delta U(x,L,T)$ of a box has been
expressed as a sum over the total energy of each normal mode. \ Before we can
discuss a maximum entropy principle involving this energy, we must know that
it is well-defined. \ We have already noted that the sum over the thermal
energy $\mathcal{U}_{T}(\omega,T)$ of the modes represents the total thermal
energy $U_{T}$ and is finite, while the sum including the zero-point energy
$\mathcal{U}_{zp}(\omega)$ is divergent. \ However, the Casimir energy $\Delta
U(x,L,T)$ in (8) can be defined as a limit and is finite. \ We recall that, in
contrast to an ideal system, any physical wave system (such as a string with
clamped ends or else electromagnetic fields in a region bounded by good
conductors) will not enforce the normal mode structure at very high
frequencies (short wavelengths). Thus it is natural to introduce a smooth
cut-off $\exp(-\Lambda\omega/c)$ related to frequency $\omega=ck$
\begin{equation}
U(L,T,\Lambda)=\Sigma_{n=1}^{\infty}\mathcal{U}(\omega_{n},T)\exp
(-\Lambda\omega_{n}/c)
\end{equation}
Next we carry out the subtractions corresponding to (8) to obtain the Casimir
energy, $\Delta U(x,L,T,\Lambda)$, and then allow the no-cut-off limit
$\Lambda\rightarrow0$. Although here we will work with an exponential cut-off
because it is easy to sum the geometric series, the result is very general;
any smooth cut-off function dependent on frequency alone will give the same result[14].

In this fashion, we can calculate the Casimir energy for the zero-point
radiation spectrum in (5),
\[
\Delta U_{zp}(x,L)=lim_{\Lambda\rightarrow0}\left\{  \Sigma_{n=1}^{\infty
}\frac{1}{2}\frac{cn\pi}{x}\exp\left(  -\Lambda\frac{n\pi}{x}\right)
+\right.
\]%
\[
\left.  +\Sigma_{n=1}^{\infty}\frac{1}{2}\frac{cn\pi}{L-x}\exp\left(
-\Lambda\frac{n\pi}{L-x}\right)  -2\Sigma_{n=1}^{\infty}\frac{1}{2}\frac
{cn\pi}{L/2}\exp\left(  -\Lambda\frac{n\pi}{L/2}\right)  \right\}
\]%
\[
=lim_{\Lambda\rightarrow0}\left\{  -\frac{c}{2}\frac{\partial}{\partial
\Lambda}\left[  \frac{1}{\exp[\Lambda\pi/x]-1}+\frac{1}{\exp[\Lambda
\pi/(L-x)]-1}-2\frac{1}{\exp[\Lambda\pi/(L/2)]-1}\right]  \right\}
\]%
\[
=lim_{\Lambda\rightarrow0}\left\{  \left[  \frac{cx}{2\Lambda^{2}\pi}%
-\frac{c\pi}{24x}+\bigcirc(\Lambda)\right]  +\left[  \frac{c(L-x)}%
{2\Lambda^{2}\pi}-\frac{c\pi}{24(L-x)}+\bigcirc(\Lambda)\right]  -\right.
\]%
\begin{equation}
\left.  -2\left[  \frac{c(L/2)}{2\Lambda^{2}\pi}-\frac{c\pi}{24(L/2)}%
+\bigcirc(\Lambda)\right]  \right\}  =-\frac{c\pi}{24}\left(  \frac{1}%
{x}+\frac{1}{L-x}-\frac{2}{L/2}\right)
\end{equation}
Thus we obtain the change in zero-point energy associated with the position
$x$ of the partition,
\begin{equation}
\Delta U_{zp}(x,L)=-\frac{c\pi}{24}\left(  \frac{1}{x}+\frac{1}{L-x}-\frac
{2}{L/2}\right)
\end{equation}
The total Casimir energy at finite temperature $T$ then involves%
\begin{equation}
\Delta U(x,L,T)=\Delta U_{T}(x,L,T)+\Delta U_{zp}(x,L)
\end{equation}
where $\Delta U$, $\Delta U_{T}$, $\Delta U_{zp}$ are formed from the
respective mode energies $\mathcal{U}$, $\mathcal{U}_{T}$, and $\mathcal{U}%
_{zp}$. \ From the result (11) for zero-point energy, we see that $\Delta
U(x,L,T)$ is finite for any spectrum $\mathcal{U}_{T}$ of thermal radiation
which has finite total energy $U_{T}$.

\section{\noindent\noindent\textbf{MAXIMUM ENTROPY PRINCIPLE FOR THERMAL
RADIATION}}

Suppose now that we consider a collection of conducting-walled boxes of length
$L$ which differ only in the placement of the partition. \ Our ideas of
uniformity suggest that the collection should have the partitions distributed
uniformly in $x$ along the open interval $(0,L)$. \ If each of these boxes
contained thermal radiation at the same temperature $T$, then they would still
differ in their total energies because of the presence of the partitions at
different locations and the associated difference in Casimir energies. \ We
notice how strikingly different this is from the situation which is usually
described in thermodynamics texts where we are dealing with ideal gas
particles in a box. \ For ideal gas particles in a box, the insertion of a
thin partition into the equilibrium situation would not change the total
energy or the entropy for the box; each box would have the same energy. \ Yet
in our situation of thermal radiation, the boxes each have the same
temperature yet each has a different energy. \ The natural situation of
maximum entropy corresponds to maximum uniformity of energy among the boxes. \ 

In order to turn this natural-maximum-entropy idea into a numerical criterion,
we must turn back to Wien's displacement theorem to note the allowed
functional form for $\mathcal{U}_{T}(\omega,T)$ and to fix the connection
between the possible radiation spectra and temperature. \ We have already
indicated that the thermodynamics associated with the Wien displacement
theorem gives thermal radiation $\mathcal{U\rightarrow U}_{RJ}(\omega)=T$
\ \ for \ \ $\omega/T<<1$. \ Thus the temperature of the random radiation
fixes the low-frequency limit of $\mathcal{U}_{T}(\omega,T)$. \ Furthermore,
the thermal energy of each normal mode must be non-negative $\mathcal{U}%
_{T}(\omega,T)\succeq0$, and the total thermal energy must be finite. \ These
criteria can be met by any distribution of normal mode thermal energies
\begin{equation}
\mathcal{U}_{T}(\omega,T)=T\,f(\omega/T)
\end{equation}
where $f$ is an arbitrary function satisfying
\begin{equation}
f(0)=1\text{, \ \ }f(x)\succeq0\text{, \ \ }\int_{0}^{\infty}dx\,f(x)<\infty
\end{equation}
If we calculate the Casimir energy $\Delta U(x,L,T)$ in (8) using (11) and
(12) with any function $f$ satisfying the criteria of (14), then the departure
$I$ from a uniform energy among the partitioned boxes is given by%
\begin{equation}
I=\sum_{i}\left\vert \Delta U(x_{i},L,T)\right\vert
\end{equation}
where the sum is over the uniformly-distributed partition locations $x_{i}$.
\ Since the Casimir energy vanishes for the partition at $x=L/2$ and is
symmetric about this point, this sum allows an immediate conversion to an
equivalent integral%
\begin{equation}
I=\int_{x=\delta}^{x=L/2}dx~|\Delta U(x,L,T)|
\end{equation}
where $\delta$ is a small cut-off distance which is much less than any other
length in the situation, $0<\delta<<min(L,c/T)$. The need for such a cut-off
arises because of the divergence of the zero-point Casimir energy at small
distances. \ The natural maximum-entropy principle states that nature will
choose the function $f(x)$ satisfying the criteria in (14) which makes the
integral $I$ in(16) a minimum.[15]

\section{\noindent\textbf{CASIMIR ENERGIES FOR VARIOUS RADIATION SPECTRA}}

In one spatial dimension, it is quick to evaluate the Casimir energies for
various monotonic spectral functions $\mathcal{U}(\omega,T)$ on a home
computer. One separates out the divergent zero-point energy contribution
corresponding to (11) and then evaluates the thermal contribution to the
Casimir energy for any assumed thermal spectrum $\mathcal{U}_{T}%
(\omega,T)=Tf(\omega/T)$ meeting the criteria of (14). \ The total Casimir
energy $\Delta U(x,L,T)$ is the sum of the thermal and zero-point
contributions. \ 

For all spectra $\mathcal{U}_{T}(\omega,T)$ satisfying the required conditions
(14), we can calculate the test integral given in Eq.(16). The spectrum
providing the smallest value for the integral appears to be the Planck
spectrum with zero-point radiation%
\begin{equation}
\mathcal{U}_{Pzp}\left(  \omega,T\right)  =\frac{1}{2}\omega\coth\left(
\frac{1}{2}\frac{\omega}{T}\right)  =\frac{\omega}{\exp(\omega/T)-1}+\frac
{1}{2}\omega
\end{equation}
Indeed, we may use the Planck form (or other functional forms) in a
variational calculation to obtain the parameters which give the smallest value
to the test integral for the given functional form. Thus the Planck form for
the thermal part of the radiation at frequency $\omega$ can be written as
\begin{equation}
\mathcal{U}_{PT}(\omega,T)=\mathcal{U}_{P}(\omega,T)-\frac{1}{2}\omega
=\frac{\omega}{\exp(\omega/T)-1}%
\end{equation}
We can introduce parameters $C_{1}$ and $C_{2}$ into a generalization of this
form giving an energy spectrum including zero-point energy as
\begin{equation}
\mathcal{U}_{C_{1}C_{2}}(\omega,T)=\frac{C_{1}\omega\exp[-C_{2}(\omega
/T)]}{1-\exp[-C_{1}(\omega/T)]}+(1/2)\omega
\end{equation}
For all positive parameters $C_{1}$ and $C_{2}$, this spectrum goes over to
energy equipartition in the limit $\omega\rightarrow0$ and goes over to
zero-point energy at high frequency while giving finite total thermal
radiation energy. Accordingly we can search for the the values of $C_{1}$ and
$C_{2}$ which make the test integral (16) a minimum. Numerical calculation
shows that the minimum value for the test integral is achieved when
$C_{1}=C_{2}=1$, corresponding to exactly the Planck spectrum (17).

Indeed, of all the functional forms tested numerically, the Planck spectrum
gave the smallest value of the test integral. We conjecture that analytic
calculation[16] would show that this spectrum provides the minimum for this
integral, and hence in this sense provides the smallest Casimir energies and
greatest uniformity in the presence of zero-point radiation.

\section{\textbf{CASIMIR ENERGY FOR THE RAYLEIGH-JEANS SPECTRUM}}

In the current textbook accounts of blackbody radiation, zero-point radiation
is not considered, and the Rayleigh-Jeans spectrum is said to be the spectrum
predicted by classical physics. \ It therefore seems of interest to calculate
the Casimir energies associated with the Rayleigh-Jeans spectrum in (5).
\ This spectrum is simple enough to allow analytic calculation of the Casimir
energies. \ We again make use of a high frequency cut-off just as in the case
of the zero-point spectrum, and calculate%
\[
\Delta U_{RJ}(x,L,T)=
\]%
\[
=lim_{\Lambda\rightarrow0}\left\{  \Sigma_{n=1}^{\infty}T\exp\left(
-\Lambda\frac{n\pi}{x}\right)  +\Sigma_{n=1}^{\infty}T\exp\left(
-\Lambda\frac{n\pi}{L-x}\right)  -2\Sigma_{n=1}^{\infty}T\exp\left(
-\Lambda\frac{n\pi}{L/2}\right)  \right\}
\]%
\[
=lim_{\Lambda\rightarrow0}\left\{  \frac{T}{\exp[\Lambda\pi/x]-1}+\frac
{T}{\exp[\Lambda\pi/(L-x)]-1}-2\frac{T}{\exp[\Lambda\pi/(L/2)]-1}\right\}
\]%
\[
=lim_{\Lambda\rightarrow0}\left\{  T\left[  \frac{x}{\Lambda\pi}-\frac{1}%
{2}+\frac{\pi\Lambda}{12x}-...\right]  +T\left[  \frac{L-x}{\Lambda\pi}%
-\frac{1}{2}+\frac{\pi\Lambda}{12(L-x)}-...\right]  +\right.
\]%
\begin{equation}
\left.  -2T\left[  \frac{L/2}{\Lambda\pi}-\frac{1}{2}+\frac{\pi\Lambda
}{12(L/2)}-...\right]  \right\}  =0
\end{equation}
Thus we find that the Rayleigh-Jeans spectrum gives no Casimir energy changes
at all. \ Indeed, the Rayleigh-Jeans spectrum is the unique spectrum which
produces no Casimir energy changes associated with the placement of the
Casimir partition, $\Delta U_{RJ}(x,L,T)=0$.[17]

\section{\noindent\textbf{'ULTRA-VIOLET CATASTROPHE' WITHOUT ZERO-POINT
RADIATION}}

We should emphasize that our maximum entropy principle indeed requires the
presence of the zero-point radiation energy. If no zero-point energy were
present, then we would still require that the thermal spectrum give energy
equipartition at low frequency and go to zero at high frequency so as to give
a finite energy density for thermal radiation. For this case, the thermal
energy would be the total energy used in (16). \ However, there would be no
natural high-frequency limit. If we tried a smooth spectrum such as that
suggested by Rayleigh $\mathcal{U}_{RT}(\omega,T)=T\exp[-C(\omega/T)]$ with an
adjustable parameter $C$ but without zero-point energy, then we would find
that the test integral given in Eq.(16) decreases as the parameter $C$
decreases, bringing the spectrum ever closer to the Rayleigh-Jeans spectrum,
in which limit the integral vanishes $I=0$ and there are no Casimir energy
changes. The absence of any natural cut-off frequency represents behavior
reminiscent of the "ultraviolet catastrophe" emphasized by Einstein and named
by Ehrenfest in 1911. What prevents the catastrophic shift of thermal
radiation to ever-higher frequencies is precisely the presence of zero-point radiation.

\section{\noindent\noindent\textbf{CONCLUDING SUMMARY }}

In this analysis, we have treated\ the thermodynamics of waves in one spatial
dimension. \ However, the ideas can be carried over to electromagnetic waves
in three spatial dimensions. \ Although the Wien displacement theorem reflects
the information from adiabatic energy changes of the known harmonic oscillator
Lagrangian for the electromagnetic modes of thermal radiation, the entropy
associated with each mode is undetermined. \ Traditional classical physics
does not find it possible to recognize a situation of natural maximum
uniformity, of maximum entropy. However, the use of Casimir energies, which
connect different radiation spectra to different total radiation energies in a
partitioned box, allows one to find a situation of natural maximum entropy.
\ \ In the absence of zero-point radiation, the entropy principle recovers
only the Rayleigh-Jeans spectrum. \ In the presence of zero-point radiation,
numerical calculation indicates that the spectrum of maximum entropy is the
Planck spectrum.

\textbf{ACKNOWLEDGEMENT }

I wish to thank Professor Daniel C. Cole for his kind invitation to present a
talk on historical aspects of the Casimir model of the electron at the Seventh
International Conference on Squeezed States and Uncertainty Relations held at
Boston University during June 2001. The beginnings of this work arose out of
communications with Professor Cole.


\begin{thebibliography}{99}                                                                                               %


\bibitem {1}See, for example, R. Eisberg and R. Resnick, \textit{Quantum
Physics of Atoms, Molecules, Solids, Nuclei, and Particles} \textit{2nd ed.}
(Wiley, New York 1985), Chapter 1.

\bibitem {2}See, for example, M. Planck, \textit{The Theory of Heat Radiation}
(Dover, New York 1959), pp. 61-63, or R. Becker and G. Leibfried,
\textit{Theory of Heat 2nd ed.} (Springer, New York 1967), pp. 16-17, or P.M.
Morse, \textit{Thermal Physics 2nd ed} (Benjamin/Cummings, Reading, MA 1969),
pp. 78-79.

\bibitem {3}See, for example, M. Planck in reference 1, pp. 72-83, or F. K.
Richtmyer, E. H. Kennard, and T. Lauritsen, \textit{Introduction to Modern
Physics} (McGraw-Hill, New York 1955), pp. 113-118, or B. H. Lavenda,
\textit{Statistical Physics: a Probabilistic Approach} (Wiley, New York 1991),
pp. 67-69. \ A very different derivation is presented by T. H. Boyer,
"Thermodynamics of the harmonic oscillator: Wien's displacement law and the
Planck spectrum," submitted for publication.

\bibitem {4}Physicists today usually regard zero-point radiation as a
"quantum" phenomenon. \ However, zero-point radiation can also be regarded as
classical random radiation, just as thermal radiation was regarded as
classical random radiation before 1900.

\bibitem {5}M. J. Sparnaay, "Measurement of the attractive forces between flat
plates," Physica \textbf{24}, 751-764 (1958); S. K. Lamoreaux, "Demonstration
of the Casimir force in the 0.6 to 6$\mu$m range," Phys. Rev. Lett.
\textbf{78}, 5-8 (1997), \textbf{81}, 5475-5476 (1998); U. Mohideen,
"Precision measurement of the Casimir force from 0.1 to 0.9 $\mu$m," Phys.
Rev. Lett. \textbf{81}, 4549-4552 (1998); and H. B. Chan, V. A. Aksyuk, R. N.
Kleiman, D. J. Bishop, and F. Capasso, "Quantum mechanical actuation of
microelectomechanical systems by the Casimir force," Science \textbf{291},
1941-1944 (2001).

\bibitem {6}H. B. G. Casimir, Proc. Kon. Ned. Akad. Wetenschap. \textbf{51},
793 (1948).

\bibitem {7}In this article we have discussed the case of waves in one spatial
dimension. However, the same thermodynamic analysis applies immediately in
three dimensions. The behavior of Casimir forces within a three-dimensional
rectangular conducting box with a conducting partition can be shown
numerically to repeat the same sort of behavior as found in the
one-dimensional case. \ Indeed, related calculations were done decades ago in
the three-dimensional calculations of M. Fierz, "Zur Anziehung leitender
Ebenen im Vacuum, "Helvetica Physica Acta \textbf{33}, 855-858 (1960), and of
T. H. Boyer, "Some Aspects of Quantum Electromagnetic Zero-Point Energy and
Retarded Dispersion Forces," Harvard doctoral thesis 1968 (unpublished),
particularly Fig. 4.

\bibitem {8}See, for example, E. A. Power, \textit{Introductory Quantum
Electrodynamics} (American Elsevier, NY 1964), pp. 18-22.

\bibitem {9}Here $\mathcal{U}$ is written in terms of the thermodynamic
potential \ $\phi(\omega/T)$ used by C. Garrod, \textit{Statistical Mechanics
and Thermodynamics} (Oxford, New York 1995), p. 128. \ See the references in 3.

\bibitem {10}See the discussion of natural units by C. Garrod, reference 9, p.
120. \ The choice $\hbar=1$ is familiar to particle physicists. \ The
measurement of temperature in energy units is familiar in thermodynamics where
our choice corresponds to the use of what is usually termed $\tau$ instead of
$T$.

\bibitem {11}See, for example, Boyer in reference 3.

\bibitem {12}It is amusing to carry out Boltzmann's derivation for the
one-dimensional case. \ We assume that the thermal energy and entropy of our
waves in a very large one-dimensional box of length $L$ satisfy $U_{T}%
(T,L)=L\,u_{T}(T),~$and $S(T,L)=L\,s(T)$ where the densities are functions of
temperature alone. \ For a normally incident plane wave, we expect a pressure
$p=\mathcal{E}/V$ rather than $p=(1/3)\mathcal{E}/V$. \ Multiplying by the
area of the walls, the force on the bounding partition corresponds to $X=u$
where $u$ is the energy per unit length. These are electromagnetic results
which involve no thermodynamics. \ Now substituting into $TdS(T,L)=dU_{T}%
(T,L)+X_{T}dL$ and separating differentials on both sides, we have
$s=2u_{T}/T$ and $ds/dT=(1/T)(du_{T}/dT)$. \ Differentiating the equation for
$s$ with respect to temperature and substituting into the second, we find a
differential equation with solution $u_{T}=\alpha T^{2}$ and so $s=2\alpha T$
where $\alpha$ is an unknown constant. \ 

\bibitem {13}H. B. G. Casimir, in reference 6, gives the force per unit area
due to electromagnetic zero-point radiation. \ The Rayleigh-Jeans spectrum
gives a different force per unit area, $F/A=-\zeta(3)k_{B}T/(4\pi x^{3})$.
\ See, for example, T. H. Boyer, "Temperature dependence of Van der Waals
forces in classical electrodynamics with classical electromagnetic zero-point
radiation," Phys. Rev. A \textbf{11}, 1650-1663 (1975).

\bibitem {14}See for example, G. H. Hardy, \textit{Divergent Series} (Oxford
University Press, London, 1956).

\bibitem {15}We notice that thermodynamics actually requires that
$\mathcal{U}(T,\omega)$ be a monotonically increasing function of frequency as
it goes from the low frequency limit $\mathcal{U}(T,\omega)=T$ to the
high-frequency limit $\mathcal{U}(T,\omega)=(1/2)\omega$. \ This already puts
additional restrictions on the monotonically decreasing functions
$f(\omega/T)$. However, we find that even monotonically decreasing functions
$f(\omega/T)$ which satisfy the limits in (14), as well as giving monotonic
functions $\mathcal{U}(T,\omega)=Tf(\omega/T)+(1/2)\omega,$ still do not
necessarily lead to monotonic Casimir energy changes $\Delta U(x,L,T)$ with
partition position $x$. \ Such functions are excluded by fundamental
thermodynamic ideas. \ However, such thermodynamic restrictions are all
subsumed by the maximum-entropy principle.

\bibitem {16}It is curious and perhaps significant that the Euler-Maclaurin
expansion which enters Casimir calculations involves the same Bernoulli
numbers as appear in the coefficients of the hyperbolic tangent function. See,
for example, R. P. Boas and C. Stutz, "Estimating sums with integrals," Am. J.
Phys. \textbf{39}, 745 (1971) and M. Abramowitz and J. Stegun, eds.,
\textit{Handbook of Mathematical Functions} (Dover, New York, 1965), pp. 804
and 806.

\bibitem {17}There are, however, Casimir forces and changes in Helmholtz free
energy. \ One also finds interesting temperature-independent entropy changes
with partition position $\Delta S_{RJ}(x,L,T)=(1/2)\left\vert \ln
[x/(L-x)]\right\vert $. \ This seems reminiscent of temperature-independent
changes associated with the mixing entropy of ideal gas particles. \ Similar
changes have been noted in quantum field theory by J. C. da Silva, A. Matos
Neto, H. Q. Placido, M. Revzen, and A. E. Santan, "Casimir effect for
conducting and permeable plates at finite temperature," Physica A
\textbf{292}, 411-421 (2001).
\end{thebibliography}
\end{document}